\documentclass[aps,pra,amsmath,amssymb,floatfix,twocolumn,amsmath,superscriptaddress,twocolumn,nofootinbib,tighten,letterpaper]{revtex4-2}
\usepackage[colorlinks,linkcolor=blue,citecolor=blue,urlcolor=blue]{hyperref}
\usepackage{multirow}
\usepackage{subfigure}
\usepackage{color}
\usepackage{mathrsfs}
\usepackage{hyperref}
\usepackage[normalem]{ulem}
\usepackage{bm}

\usepackage{amssymb}   % for math
\usepackage{amsmath}
\renewcommand\vec[1]{\ensuremath\boldsymbol{#1}} % bold font for vectors

\usepackage{amsfonts, relsize, color}
\usepackage{graphics}
\usepackage{graphicx}
\usepackage{subfigure}
\usepackage{hyperref}
\usepackage{color}
\usepackage{comment}

\newcommand{\bra}[1]{\left< #1 \right|}
\newcommand{\ket}[1]{\left| #1 \right>}

\let\phi=\varphi
\let\epsilon=\varepsilon

\renewcommand{\vec}[1]{\bm{#1}}

\definecolor{DarkRed}{rgb}{0.80,0,0}
\definecolor{DarkGray}{rgb}{0.7,0.7,0.7}
\newcommand{\prlsection}[1]{\textit{#1}.\kern0.05em---\kern0.05em\ignorespaces}
\begin{document}
\title{Zero modes and index theorems for non-Hermitian Dirac fermions}

\author{Bitan Roy}
\affiliation{Department of Physics, Lehigh University, Bethlehem, Pennsylvania, 18015, USA}
\affiliation{Centre for Condensed Matter Theory, Department of Physics, Indian Institute of Science, Bengaluru 560012, India}

\date{\today}

%%%%%%%%%%%%%%%%%%%%%%%%%%%%%%%%%%%%%%%%%%%%%%%%%%%%%%%%%%%%%%%%%%%%%%%%%%%%%%
%%%%%%%%%%%%%%%%%%%%%%%%%%%%%%%%%%%%%%%%%%%%%%%%%%%%%%%%%%%%%%%%%%%%%%%%%%%%%%
\begin{abstract}
Dirac fermions, subject to external magnetic fields and in the presence of mass orders that assume topologically nontrivial spatial textures such as domain wall and vortices, for example, bind robust midgap states at zero energy, the number of which is governed by the Aharonov-Casher and Jackiw-Rebbi or Jackiw-Rossi index theorems, respectively. Here I extend the jurisdiction of these prominent index theorems to Lorentz invariant non-Hermitian (NH) Dirac operators, constructed by augmenting the celebrated Dirac Hamiltonian by a masslike anti-Hermitian operator that also scales linearly with momentum. The resulting NH Dirac operator manifests all-real eigenvalues over an extended NH parameter regime, characterized by a real effective Fermi velocity for NH Dirac fermions ($v_{_{\rm F}}$). From the explicit solutions of the zero-energy bound states, I show that in the presence of external magnetic fields of arbitrary shape such modes always exist when the system encloses a finite number of magnetic flux quanta, while in the presence of spatially nontrivial textures of the mass orders localized zero-energy modes can only be found in the spectrum when $v_{_{\rm F}}$ is real. These findings pave a concrete route to realize nucleation of competing orders from the topologically robust zero-energy manifold in NH or open Dirac systems. Possible extensions of these outcomes to other index theorems and tabletop experimental setups to test these predictions are discussed.      
\end{abstract}
%%%%%%%%%%%%%%%%%%%%%%%%%%%%%%%%%%%%%%%%%%%%%%%%%%%%%%%%%%%%%%%%%%%%%%%%%%%%%%
%%%%%%%%%%%%%%%%%%%%%%%%%%%%%%%%%%%%%%%%%%%%%%%%%%%%%%%%%%%%%%%%%%%%%%%%%%%%%%
\maketitle

%-------------------------------------------------------------------------------%
%                                 MAIN ARTICLE                                  %
%-------------------------------------------------------------------------------%

\section{Introduction}

Zero energy in the Dirac theory enjoys a special status as the associated eigenvalue spectrum extends equally to positive and negative values~\cite{Dirac:1, Dirac:2, Dirac:3}. Therefore, any bound state that resides precisely at zero energy often (if not always) is robust due to the intrinsic particle-hole symmetry in the system. Such a situation occurs at least in two specific cases. (a) When massless Dirac fermions, confined to an Euclidean plane, are subject to external uniform or inhomogeneous magnetic fields, the number of zero-energy bound states is exactly equal to the total magnetic flux quanta enclosed by the system, guaranteed by the Aharonov-Casher index theorem~\cite{aharonovcasher:1, aharonovcasher:2, aharonovcasher:3}. (b) Bound states at zero energy appear in the spectrum of Dirac fermions when they foster mass orders, captured by Hermitian operators fully anticommuting with the Dirac Hamiltonian, that assume spatially nontrivial topological textures. For example, in one spatial dimension, when the Dirac mass takes the shape of a domain wall a zero-energy mode gets pinned where the mass order changes sign, a result known as the Jackiw-Rebbi index theorem~\cite{jackiwrebbi:1, jackiwrebbi:2, jackiwrebbi:3}. In two spatial dimensions, when a composite Dirac mass, described by two mutually anticommuting mass matrices with the requisite U(1) symmetry between them, takes the texture of a vortex, $n$ number of zero-energy bound states get pinned near the core of such topological defects, where $n$ is the vorticity of the mass texture. Such a one-to-one correspondence between the vorticity of the complex mass order and the number of zero-energy bound states goes by the name of the Jackiw-Rossi index theorem~\cite{jackiwrossi:1, jackiwrossi:2}.

The above-mentioned three index theorems, although originated in the context of high-energy physics, found their direct and important implications in various quantum materials, featuring emergent Dirac quasiparticles (massless or massive) as low-energy excitations around a few isolated points in the Brillouin zone, known as the Dirac or Weyl materials~\cite{DiracMat:1, DiracMat:2, DiracMat:3}. For example, the Jackiw-Rebbi index theorem directly applies to the Su-Schrieffer–Heeger chain in one dimension, when the effective Dirac Hamiltonian features a domain-wall mass~\cite{SSH:1, SSH:2}. The same mechanism is also responsible for the existence of zero-energy one-dimensional edge modes and two-dimensional surface states at the boundaries of two- and three-dimensional strong topological insulators, respectively~\cite{TIRMP:1, TIRMP:2}, and zero-energy flat band in uniaxially strained honeycomb lattice~\cite{uniaxialstrain:LL}. The Aharonov-Casher index theorem, on the other hand, dictates the robustness of the zeroth Landau level for graphene, subject to external magnetic fields of arbitrary spatial profile~\cite{aharonovcasher:4, aharonovcasher:5}, which is applicable to Dirac fermions subject to chiral or axial magnetic fields~\cite{aharonovcasher:6, aharonovcasher:7, aharonovcasher:8} as well as the ones residing on a negatively curved hyperbolic space~\cite{aharonovcasher:9}. Finally, the signature of the Jackiw-Rossi index theorem gets imprinted on the zero-energy modes near the vortex core of an $s$-wave superconductor otherwise realized on the surface of a three-dimensional topological insulator via proximity effect~\cite{JaRoEx:1}. The same index theorem also applies to graphene when the ground state harbors a Kekul\'e valence-bond order~\cite{JaRoEx:2}, assisted by electron-phonon interaction~\cite{JaRoEx:3} or an easy-plane antiferromagnetic order, assisted by on-site Hubbard repulsion and in-plane Zeeman coupling~\cite{JaRoEx:4}. In both cases, the mass orders possess the requisite U(1) symmetry in the ordered states and zero-energy modes are found in the spectrum. Existence of such zero-energy modes leads to fascinating phenomena such as charge fractionalization~\cite{JaRoEx:5, JaRoEx:6, JaRoEx:7, JaRoEx:8} and competing orders near the vortex core~\cite{JaRoEx:9, JaRoEx:10, JaRoEx:11}.

In this work, I set out to establish the footprints of these three key index theorems on a class of quasirelativistic open quantum systems, effectively described by non-Hermitian (NH) Dirac operators. In this pursuit, I strictly focus on a specific family of NH Dirac operators that enjoy all the quintessential elegance of the original Dirac theory~\cite{Dirac:1, Dirac:2, Dirac:3}. Namely, such NH Dirac operators support guaranteed all-real eigenvalues over an extended NH parameter regime and display quasirelativistic Lorentz invariance. Therefore, an introduction to the construction of such a NH Dirac theory would greatly facilitate the forthcoming summary of the central outcomes of this work, which I present in Sec.~\ref{SummaryIntro}.

\subsection{NH Dirac theory: Construction}~\label{subsec:construction}

The Hamiltonian for massless Dirac fermions in $d$ spatial dimensions takes the following universal form~\cite{Dirac:1, Dirac:2, Dirac:3} 
\begin{equation}~\label{eq:DiracHamil}
H_{\rm Dir}(\vec{k})= v_{_{\rm H}} \sum^{d}_{j=1} \; \left( \Gamma_j k_j \right),
\end{equation}
where $v_{_{\rm H}}$ is a real parameter, bearing the dimensionality of the Fermi velocity, $k_j$s are the Cartesian components of the spatial momentum $\vec{k}=(k_1,\cdots, k_d)$, and $\Gamma_j$s are mutually anticommuting Hermitian matrices satisfying the Clifford algebra $\{ \Gamma_j, \Gamma_k \}=2 \delta_{jk}$ where $j,k=1,\cdots,d$ and $\delta_{jk}$ is the Kronecker delta symbol. The dimensionality of the $\Gamma$ matrices and that of the associated Dirac spinors depend on the microscopic details, which I do not delve into at this time. Due to the unique Fermi velocity in all directions, $H_{\rm Dir}(\vec{k})$ is invariant under the Lorentz transformation or equivalently it transforms as a \emph{scalar} under the Lorentz transformation. The energy spectrum of the Dirac Hamiltonian is composed of two branches $\pm E(\vec{k})$, where $E(\vec{k})=v_{_{\rm H}}|\vec{k}|$ manifests the linear energy-momentum relationship, the hallmark of massless relativistic fermions with the Fermi velocity $v_{_{\rm H}}$ playing the role of the speed of light ($c$). Typically, in relativistic quantum crystals $v_{_{\rm H}} \ll c$.

The Dirac theory permits another class of Lorentz scalars, namely the Dirac masses, represented by a set of Hermitian matrices $\{ M \}$ that fully anticommutes with $H_{\rm Dir}(\vec{k})$ and squares to the identity matrix~\cite{peskin}. In terms of these two Lorentz invariant quantities, I define an anti-Hermitian operator $M H_{\rm Dir}(\vec{k})$, which also transforms as a scalar under the Lorentz transformation. The Lorentz invariant NH Dirac operator is then given by~\cite{NHDirac:1, NHDirac:2, NHDirac:3, NHDirac:4, NHDirac:5}  
\begin{equation}~\label{eq:NHDirac}
H^{\rm NH}_{\rm Dir} (\vec{k})= v_{_{\rm H}} \sum^{d}_{j=1} \; \left( \Gamma_j k_j \right) - v_{_{\rm NH}}\;  \left\{ M \; \sum^{d}_{j=1} \; \left( \Gamma_j k_j \right) \right\},
\end{equation}
where $v_{_{\rm NH}}$ is a real parameter, also bearing the dimension of the Fermi velocity. The eigenspectrum of $H^{\rm NH}_{\rm Dir}(\vec{k})$ is also composed of two branches $\pm E_{\rm NH}(\vec{k})$ with $E_{\rm NH}(\vec{k})=\sqrt{v^2_{_{\rm H}}-v^2_{_{\rm NH}}} \; |\vec{k}|$ that continues to feature the signature linear energy-momentum relation for nodal relativistic quasiparticles. The quantity $v_{_{\rm F}}=\sqrt{v^2_{_{\rm H}}-v^2_{_{\rm NH}}}$ is the effective Fermi velocity of NH Dirac fermions, which now plays the role of the speed of light in open Dirac systems. In what follows, I define a dimensionless quantity $\alpha=v_{_{\rm NH}}/v_{_{\rm H}}$ and conveniently set $v_{_{\rm H}}=1$. When $|\alpha| < 1$ the eigenspectrum of $H^{\rm NH}_{\rm Dir}(\vec{k})$ and $v_{_{\rm F}}$ are purely real, whereas for $|\alpha|>1$ all the eigenvalues of $H^{\rm NH}_{\rm Dir}(\vec{k})$ and $v_{_{\rm F}}$ are purely imaginary. For $\alpha=\pm 1$, all the eigenvalues of $H^{\rm NH}_{\rm Dir}(\vec{k})$ and $v_{_{\rm F}}$ are equal to zero, which mark the exceptional points in this construction. Throughout this work, I stay away from such singular points.

In this work, I scrutinize the Aharonov-Casher, Jackiw-Rebbi, and Jackiw-Rossi index theorems within the framework of such NH Dirac operators. An important comment is due at this stage. In the last two cases, the Hermitian matrices describing the mass orders that assume spatially nontrivial topological textures (domain wall and vortices) are described by Hermitian operators that \emph{commute} with $M$ from Eq.~\eqref{eq:NHDirac} and anticommute with $H_{\rm Dir}(\vec{k})$ from Eq.~\eqref{eq:DiracHamil}, such that they fully anticommute with anti-Hermitian operator $M H_{\rm Dir}(\vec{k})$ and thus anticommute with the total NH operator $H^{\rm NH}_{\rm Dir}(\vec{k})$, hence representing genuine mass orders for NH Dirac fermions. Only then the eigenspectrum of the corresponding total NH massive Dirac operator can be purely real over an extended NH parameter regime, as then the effective single-particle Hamiltonian in the ordered phase is captured by matrix operators all of which mutually anticommute with each other. Such mass ordering in the context of NH Dirac theory is named ``commuting class masses''~\cite{NHDirac:1}, which will be discussed in Secs.~\ref{sec:JackiwRebbi} and~\ref{sec:JackiwRossi}. Next, I present a synopsis of the main results.

\subsection{Summary of main results}~\label{SummaryIntro}

I begin the discussion by scrutinizing the fate of the Aharonov-Casher index theorem for planar massless NH Dirac fermions subject to external perpendicular magnetic fields of arbitrary shape such that the system encloses $N$ number of magnetic flux quanta. Such a system is shown to host exactly $N$ number of normalizable zero-energy modes irrespective of the value of the NH parameter ($\alpha$) as their localization length is solely determined by the magnetic field profile and is completely insensitive to the effective Fermi velocity of NH Dirac fermions ($v_{_{\rm F}}$). See Sec.~\ref{sec:AharonovCasher} for details.

Next, I establish the Jackiw-Rebbi index theorem for a one-dimensional massive NH Dirac system in which the mass operator that assumes a domain-wall texture \emph{commutes} with the mass matrix $M$ entering the construction of the NH Dirac operator in Eq.~\eqref{eq:NHDirac}. Such mass orders are named ``commuting class masses.'' In such a system, I find localized and normalizable zero-energy modes at the core of the domain wall only when the effective Fermi velocity of the collection of NH Dirac fermions is real. This outcome can be justified in the following way. The ratio of the magnitude of the asymptotic value of the mass as $x \to \pm \infty$ to the effective Fermi velocity of NH Dirac fermions determines the localization length of the zero-energy modes bound to the domain-wall defect core. Naturally, only when the effective Fermi velocity is real the localization length of zero modes is also real and they can be found in the spectrum of massive NH one-dimensional Dirac fermions. See Sec.~\ref{sec:JackiwRebbi} for details.

Finally, I extend the jurisdiction of the Jackiw-Rossi index theorem for two-dimensional massive Dirac fermions in the presence of a composite mass order with the U(1) symmetry that assumes a vortex-type real space topological defect in a NH setup. The mass matrix $M$ in the construction of the NH Dirac operator [Eq.~\eqref{eq:NHDirac}] is chosen such that it commutes with the composite mass order, which thus once again belongs to the family of ``commuting class masses.'' In this setup, I show that a mass vortex of integer vorticity $n$ supports exactly $n$ number of zero-energy modes only when the effective Fermi velocity of NH Dirac fermions is real. By exploiting a pseudo-particle-hole symmetry of the NH Jackiw-Rossi operator, I extend this model to include additional terms therein yielding the generalized NH Jackiw-Rossi operator, that, for example include the orbital coupling of a gauge field required for the stability of isolated or deconfined vortices. In this construction, I show that the system supports only one zero-energy mode when $n$ is an odd integer, while the spectrum is devoid of any such zero modes for any even $n$. Thus, in terms of the number of zero-energy modes, the generalized NH Jackiw-Rossi operator features a $Z_2$ index protected by a pseudo-particle-hole symmetry, while the conventional NH Jackiw-Rossi operator renders a $Z$ index, protected by a unitary particle-hole symmetry. See Sec.~\ref{sec:JackiwRossi} for details. I also note that when the Fermi velocity of NH Dirac fermions becomes purely imaginary, i.e. when $|\alpha|>1$, these zero-energy bound states tied to domain-wall and vortices become plane waves at zero-energy, a phenomenon that only occurs in NH Dirac systems.

Although in this work I focus on three prominent index theorems for quasirelativistic Dirac fermions in open or NH setups, based on the outcomes one can conjecture the following generic statement. The spectrum of Lorentz invariant NH Dirac fermions should support the same number of zero-energy modes as in Hermitian systems for arbitrary strength of the NH parameter if the localization length of such midgap bound states is independent of the effective Fermi velocity, as is the case for the Aharonov-Casher index theorem for gapless or massless Dirac fermions. On the other hand, if such bound states arise in the spectrum of massive Dirac fermions, then their number remains the same as in the Hermitian system, but only if the effective Fermi velocity is real, as is the case for Jackiw-Rebbi and Jackiw-Rossi index theorems. Finally, it should be noted that for massive Dirac fermions such index theorems hold in NH systems when the mass orders, assuming topologically nontrivial spatial textures (such as domain wall and vortices, for example), belong to the ``commuting class mass'' family.    

\subsection{Organization}

The remainder of the discussion is organized as follows. In Sec.~\ref{sec:AharonovCasher}, I show the Aharonov-Casher index theorem for planar NH gapless Dirac fermions, subject to external perpendicular magnetic fields. Section~\ref{sec:JackiwRebbi} is devoted to demonstrating the Jackiw-Rebbi index theorem for NH one-dimensional massive Dirac fermions in the presence of a domain-wall of a ``commuting class mass.'' The Jackiw-Rossi index theorem of a composite mass order for two-dimensional NH Dirac systems is discussed in Sec.~\ref{sec:JackiwRossi}. A summary of results, their possible extensions, and promising tabletop experimental platforms to test the theoretical predictions are staged in Sec.~\ref{sec:discussion}.

\section{Aharonov-Casher index theorem}~\label{sec:AharonovCasher}

The Aharonov-Casher index theorem concerns the number of zero-energy bound states when planar massless Dirac fermions experience uniform or nonuniform perpendicular magnetic fields, such that the system encloses $N$ number of magnetic flux quanta. In this Section, I establish this index theorem for a collection of NH massless Dirac fermions, described by the NH operator from Eq.~\eqref{eq:NHDirac}. For concreteness, I consider its minimal two-component representation in $d=2$, for which $\Gamma_1=\sigma_1$ and $\Gamma_2=\sigma_2$. Then a natural and unique choice of $M$ is $M=\sigma_3$. Here, $\{ \sigma_\mu \}$ is the set of two-dimensional Pauli matrices with $\mu=1,2,3$. The explicit form of the NH operator in the presence of an external magnetic field reads   
\begin{equation}
H^{\rm NH}_{\rm Dir}(\vec{k} \to -i {\boldsymbol \nabla}, \alpha, \vec{A})= {\boldsymbol \sigma}_{_\perp} \cdot {\boldsymbol \Pi}_\perp +  \alpha \sigma_3 \left({\boldsymbol \sigma}_{_\perp} \cdot {\boldsymbol \Pi}_\perp \right), 
\end{equation}
where ${\boldsymbol \Pi}_\perp=(\Pi_1, \Pi_2)$ with $\Pi_j=-i \partial_j-e A_j$ for $j=1,2$ and ${\boldsymbol \sigma}_{_\perp}=(\sigma_1, \sigma_2)$. Here $\vec{A}$ is the magnetic vector potential and $e$ is the electronic charge. As in the original work by Aharonov and Casher~\cite{aharonovcasher:1}, I work in the Coulomb gauge such that ${\boldsymbol \nabla} \cdot \vec{A}=0$, which can be satisfied with the choice of $\vec{A}=(-\partial_y \chi, \partial_x \chi)$, where $\chi \equiv \chi(\vec{r})$ is a scalar function. In terms of the scalar function $\chi(\vec{r})$, the magnetic field ($\vec{B}$) is given by $\vec{B} = {\boldsymbol \nabla} \times \vec{A}=\nabla^2 \chi(\vec{r}) \hat{z}$, where $\hat{z}$ is the unit vector in the $z$ direction.

Here I show the explicit solutions for the right eigenvectors $\ket{\Psi_0}_R$ associated with the zero-energy modes, satisfying the secular equation
\begin{equation}
H^{\rm NH}_{\rm Dir}(\vec{k} \to -i {\boldsymbol \nabla}, \alpha, \vec{A}) \; \ket{\Psi_0}_R =0.
\end{equation}
In the matrix notation, $\ket{\Psi_0}_R$ is given by a two-component column vector with $\ket{\Psi_0}_R=(\Psi_{+,R}, \Psi_{-,R})^\top$. Therefore, in terms of the components of $\ket{\Psi_0}_R$, the above equation for the right eigenvectors associated with the zero-energy modes takes the form 
\begin{equation}
(1-\alpha \; \kappa) \left(\Pi_1 + i \kappa \Pi_2 \right) \Psi_{\kappa, R} =0
\end{equation}
for $\kappa=\pm$. It is worth noticing that the NH parameter ($\alpha$) scales out of the differential operators that act on the spinor $\ket{\Psi_0}_R$. As a result, $\alpha$ does not enter the explicit solutions for the zero-energy modes. The explicit solutions for the zero modes can now be found following the original work of Aharonov and Casher, which is worth discussing here for the sake of completeness. Notice that 
\begin{equation}
\{ H^{\rm NH}_{\rm Dir}(\vec{k} \to -i {\boldsymbol \nabla}, \alpha, \vec{A}), \sigma_3 \}=0
\end{equation}
and therefore the zero modes are also eigenstates of the diagonal Pauli matrix $\sigma_3$, the generator of the unitary particle-hole symmetry, and explicitly it is given by   
\begin{eqnarray}~\label{eq:rightzeroAhaCash}
\ket{\Psi_0}_R = C \; |\vec{r}|^{j} \; e^{i (j\phi)} \; \exp[-e \chi(\vec{r})] \left( \begin{array}{c} 
1 \\
0
\end{array}
\right), 
\end{eqnarray}
where $\phi$ is the polar angle, $C$ is the normalization constant, discussed in a moment, and $j$ is an integer required for the single-valued right wave functions. The scalar function $\chi(\vec{r})$ is obtained from the spatial profile of the external magnetic field $B(\vec{r})$ via the Green's function associated with the Laplacian in two dimensions as  
\begin{equation}
\chi(\vec{r})= \int \frac{d^2 \vec{r}^\prime}{2 \pi} \; \ln|\vec{r}^\prime - \vec{r}| \; B(\vec{r}^\prime).
\end{equation}
Therefore, in the $|\vec{r}| \to \infty$ limit 
\begin{align}
\exp[-e \chi(\vec{r})] \to \exp\left[ e \ln|\vec{r}| \int \frac{d^2 \vec{r}^\prime}{2 \pi} B(\vec{r}^\prime) \right]
=  r^{-\frac{e \Phi}{2 \pi}},
\end{align}
where $\Phi =\int d^2\vec{r}^\prime B(\vec{r}^\prime)$ is the total magnetic flux enclosed by the system. Notice that in the natural units ($\hbar=1$) the magnetic flux quantum $\Phi_0=h/e \to 2\pi/e$ and thus $e \Phi/2\pi \equiv \Phi/\Phi_0=N$ is the total magnetic flux quanta enclosed by the system.

The left eigenvectors for the zero-energy modes, given by the row vector ${_L}\bra{\Psi_0}=(\Psi_{L,+}, \Psi_{L,-})$ for the zero-energy modes are found from the solutions of 
\begin{eqnarray}
{_L}\bra{\Psi_0} \; H^{\rm NH}_{\rm Dir}(\vec{k} \to -i {\boldsymbol \nabla},\alpha, \vec{A}) = 0.
\end{eqnarray}
Once again the NH parameter ($\alpha$) does not enter the solution for the left eigenvectors for the zero modes, for which I find 
\begin{equation}~\label{eq:leftzeroAhaCash}
{_L}\bra{\Psi_0} = \left( \ket{\Psi_0}_R \right)^\dagger
= C^\ast |\vec{r}|^{j} \; e^{i (j\phi)} \; \exp[-e \chi(\vec{r})] \left( \begin{array}{cc}
1 & 0
\end{array}
\right).
\end{equation}
The normalization condition is then fixed by the biorthonormal condition, given by~\cite{biorthoQM}  
\begin{equation}~\label{eq:biorthonormality}
\int \frac{d^2 \vec{r}}{(2\pi)^2} \; {_L}\bra{\Psi_0} \Psi_0\rangle_R =1, 
\end{equation}
as I am dealing with the eigenstates of a NH operator, from which one can immediately determine the normalization constant $C$. However, for the normalizable solutions for the zero-energy mode there exists a constraint between $j$ and $N$, given by 
\begin{equation}
1+ 2 j - 2 N <0
\end{equation}
such that the integrand in the above normalization condition goes to zero as $r \to \infty$. This condition imposes a restriction on the allowed integer values of $j$, given by 
\begin{equation}
j=0,1, \cdots, N-1/2
\end{equation}
for which the solutions remain normalizable. Notice that the $N$ allowed values of $j$ are completely independent of the profile of the magnetic field, and are solely determined by the total magnetic flux quanta $N$ enclosed by the planar NH quasirelativistic system. Furthermore, these solutions can always be found for any arbitrary value of $\alpha$ or the effective Fermi velocity $v_{_{\rm F}}$ of the collection of NH Dirac fermions. This is so because the localization length of the zero modes is solely determined by the magnetic flux profile in which $v_{_{\rm F}}$ plays no role, see Eqs.~\eqref{eq:rightzeroAhaCash} and~\eqref{eq:leftzeroAhaCash}. Hence, the Aharonov-Casher index theorem applies to quasirelativistic massless NH Dirac fermions in two dimensions, irrespective of the strength of non-Hermiticity in the system. Next, I discuss two cases where the reality condition of the effective Fermi velocity of NH Dirac fermions plays a decisive role for the existence of normalizable zero-energy modes.

\section{Jackiw-Rebbi index theorem}~\label{sec:JackiwRebbi}

The Jackiw-Rebbi index theorem applies to one-dimensional massive Dirac fermions for which the mass order $m(x)$, accompanied by a Hermitian matrix that fully anticommutes with the free Dirac Hamiltonian assumes the profile of a domain wall, given by $m \; (x \to \pm \infty)=\pm m_0$ (constant) otherwise arbitrary~\cite{jackiwrebbi:1}. In one-dimensional NH Dirac systems, the corresponding NH Dirac operator can be obtained by taking $\Gamma_1=\sigma_1$ and $M=\sigma_3$, and the mass order is represented by the Hermitian operator $m(x) \sigma_3$. The corresponding total massive NH Dirac operator then reads as 
\begin{equation}~\label{eq:NHdomainwall}
H^{\rm NH}_{\rm Dir} (\vec{k} \to -i {\boldsymbol \nabla}, \alpha)= \left( \sigma_1 + i \alpha \sigma_2 \right) \left( -i \partial_x \right) + m(x) \sigma_3.   
\end{equation}
Before I delve into the search for the zero modes in the spectrum of the above NH operator, it is worth pausing to discuss the choice of the Hermitian matrix for the mass order. For this purpose and without any loss of generality, I ignore any spatial modulation of the mass order and assume it to be constant or uniform $m(x)=m_0$ for any $x$. The one-dimensional Dirac Hamiltonian $\sigma_1 k_x$ supports two mass matrices, namely $\sigma_2$ and $\sigma_3$. Once I choose $M=\sigma_3$ in Eq.~\eqref{eq:NHDirac} to construct the NH Dirac operator, only for one mass matrix $\sigma_3$ one can find a real definite eigenvalue spectrum for $|\alpha|<1$ of the corresponding total NH massive Dirac operator given by $(\sigma_1+i \alpha \sigma_2)k_x + \sigma_3 m_0$. Explicitly, the eigenvalues of this NH operator are given by $\pm \sqrt{(1-\alpha^2) k^2_x +m^2_0}$. Such mass order is called ``commuting class mass'' as $[M,\sigma_3]=0$~\cite{NHDirac:1}. In this Section and the next one, I only focus on such ``commuting class mass'' orders.

Now I return to finding the zero-energy bound state in the spectrum of the NH operator shown in Eq.~\eqref{eq:NHdomainwall}. In the Hermitian system, the corresponding operator satisfies a unitary particle-hole symmetry, generated by $\sigma_2$ as $\{ H^{\rm NH}_{\rm Dir} (\vec{k} \to -i {\boldsymbol \nabla}, 0), \sigma_2 \}=0$. Hence, any mode at precise zero energy is an eigenstate of $\sigma_2$. However, in the NH setup $\sigma_2$ generates the \emph{pseudo}-particle-hole symmetry~\cite{pseudoPH:1, pseudoPH:2}, given by 
\begin{eqnarray}
\sigma_2 H^{\rm NH}_{\rm Dir} (\vec{k} \to -i {\boldsymbol \nabla}, \alpha) \sigma_2=-\left[ H^{\rm NH}_{\rm Dir} (\vec{k} \to -i {\boldsymbol \nabla}, \alpha) \right]^\dagger. \nonumber \\
\end{eqnarray}
The pseudo-particle-hole symmetry guarantees that if there exists a right eigenvector for the zero-energy mode then there must also exist its dual the left eigenvector. With this symmetry in hand I proceed to find the right zero-energy eigenvector $\ket{\Psi_0}_R = (u_R,  v_R)^\top$ in the spectrum of NH massive Dirac fermions with a domain-wall mass by solving the following differential equation
\begin{equation}
\left[ \left( \sigma_1 + i \alpha \sigma_2 \right) \left( -i \partial_x \right) + m(x) \sigma_3 \right]
\left(\begin{array}{c}
u_R \\
v_R 
\end{array} \right) =0.
\end{equation}
After some straightforward algebra, I find the zero-energy right eigenmode bound to the domain-wall defect for the mass order, which is given by 
\begin{equation}
\ket{\Psi_0}_R (x,\alpha)= C \exp\left[ - \int^{x}_0 dx^\prime \; \frac{m(x^\prime)}{\sqrt{1-\alpha^2}} \right] \; 
\left(\begin{array}{c}
1 \\
i 
\end{array} \right)
\end{equation}
and it is fascinatingly an eigenstate of $\sigma_2$, where $C$ is the normalization constant. Notice that this solution yields a localized bound state only when $|\alpha|<1$ or equivalently the effective Fermi velocity of NH Dirac fermions is real or when all the eigenvalues of the corresponding NH operator are also purely real. Exploiting the pseudo-particle-hole symmetry I can immediately arrive at the left zero-energy eigenmode, given by ${_L}\bra{\Psi_0}(x,\alpha)=\left( \ket{\Psi_0}_R (x,-\alpha) \right)^\dagger$. The normalization constant $C$ can readily be obtained from the biorthonormality condition, shown in Eq.~\eqref{eq:biorthonormality}. Therefore, the Jackiw-Rebbi index theorem applies to the ``commuting class mass'' domain wall in pseudorelativistic NH Dirac systems in one dimension and it continues to host localized zero-energy left and right eigenmodes as long as the Fermi velocity of NH Dirac fermions is real. Finally, with a specific form of the domain-wall mass $m(x)=m_0 \tanh(x)$~\cite{Stojkovic:1}, I find 
\begin{equation}
\ket{\Psi_0}_R (x,\alpha)= C \; \left[ {\rm sech}(x) \right]^{m_0/\sqrt{1-\alpha^2}} \; 
\left(\begin{array}{c}
1 \\
i 
\end{array} \right).
\end{equation}
Notice that the bound state in the presence of a domain wall for $|\alpha|<1$ becomes a plane-wave solution at zero energy for  $|\alpha|>1$. The existence of such a zero-energy plane-wave solution in the presence of an underlying domain wall is only possible in NH Dirac systems. Next, I proceed to scrutinize the Jackiw-Rossi index theorem in two dimensions in NH Dirac systems.

\section{Jackiw-Rossi index theorem}~\label{sec:JackiwRossi}

The Jackiw-Rossi index theorem dictates the number of zero-energy bound states in the spectrum of two-dimensional massive Dirac fermions, when two mutually anticommuting Dirac masses with respective amplitudes of $\Delta_1$ and $\Delta_2$ and a requisite U(1) symmetry between them feature vortexlike topological defects in real space. Here, I extend this index theorem for NH Dirac fermions for which the corresponding operator is constructed following the general principle from Eq.~\eqref{eq:NHDirac}. Before delving into such real space topological defects for the composite (two-component) mass order, consider the corresponding NH operator in the presence of two uniform masses that will allow us to uniquely identify the mass matrix $M$, appearing in the anti-Hermitian component of the NH Dirac operator in Eq.~\eqref{eq:NHDirac}, given by
\begin{equation}~\label{eq:JRuniform}
H^{\rm NH}_{\rm Dir} (\vec{k},\alpha)= (1-\alpha M) \sum_{j=1,2} \Gamma_j k_j + \sum_{j=1,2} \Delta_j \Gamma_{2+j}.
\end{equation}
First consider the Hermitian limit ($\alpha=0$) of this operator. The massive Dirac Hamiltonian then involves four mutually anticommuting Hermitian matrices  $\Gamma_j$ with $j=1,\cdots, 4$ each of which squares to the identity matrix. Therefore, the minimal dimensionality of the $\Gamma$ matrices in this case must be four. On the other hand, the maximal number of mutually anticommuting four-dimensional Hermitian $\Gamma$ matrices is five, constituting the set $\{\Gamma_j \}$ where $j=1,\cdots 5$. As all representations of five mutually anticommuting four-dimensional Hermitian $\Gamma$ matrices are unitarily equivalent~\cite{Clifford}, without any loss of generality, I here work with their following explicit representation
\begin{eqnarray}~\label{eq:gammarepresentation}
\Gamma_1 &=& \tau_3 \otimes  \sigma_1, \:
\Gamma_2 = \tau_3 \otimes  \sigma_2, \:
\Gamma_3 = \tau_1 \otimes  \sigma_0, \nonumber \\
\Gamma_4 &=& \tau_2 \otimes  \sigma_0, \: \text{and} \:
\Gamma_5 = \tau_3 \otimes  \sigma_3
\end{eqnarray}
for the sake of convenience, where $\otimes$ corresponds to the Kronecker product. Four of these $\Gamma$ matrices appear in the massive Dirac Hamiltonian and thus the fifth member of this set $\Gamma_5$ \emph{anticommutes} with the Hamiltonian and generates its unitary particle-hole symmetry.

The Dirac Hamiltonian for free fermions in two dimensions involves only \emph{two} mutually anticommuting $\Gamma$ matrices, namely $\Gamma_1$ and $\Gamma_2$, and the theory enjoys an SU(2) chiral symmetry, generated by $\{ \Gamma_{34}, \Gamma_{45}, \Gamma_{53} \}$, where $\Gamma_{jk}=i \Gamma_j \Gamma_k$~\cite{graphenemass:1, graphenemass:2}. The two-dimensional four-component Dirac system altogether supports \emph{four} mass orders. Three of them break the SU(2) chiral symmetry and constitute the set of chiral symmetry breaking mass orders, explicitly given by $\{\Gamma_3, \Gamma_4, \Gamma_5 \}$. Specifically in two dimensions there exists a fourth mass order, represented by $\Gamma_{12}$, which transforms as a scalar under the chiral rotation. I name it a ``chiral scalar'' mass, which, however, breaks the time-reversal symmetry that in the announced representation in Eq.~\eqref{eq:gammarepresentation} is generated by the antiunitary operator ${\mathcal T}= \Gamma_{14} {\mathcal K}$, where ${\mathcal K}$ is the complex conjugation. The chiral symmetry breaking mass orders preserve the time-reversal symmetry. While all the four mass orders anticommute with the free Dirac Hamiltonian, the chiral scalar mass operator \emph{commutes} with all the members of the chiral symmetry breaking mass order. This observation leads to a \emph{unique} choice of $M=\Gamma_{12}$ in Eq.~\eqref{eq:JRuniform} in the construction of the NH Dirac operator such that two mass order matrices $\Gamma_3$ and $\Gamma_4$ belong to the ``commuting class mass'' family for NH Dirac fermions. The eigenspectrum of the total NH massive Dirac operator is the given by $\pm E_{\alpha}(\vec{k})$, where
\begin{equation}
E_{\alpha}(\vec{k})=\sqrt{(1-\alpha^2) |\vec{k}|^2 + \Delta^2_1 + \Delta^2_2}.
\end{equation}
Notice that $E_{\alpha}(\vec{k})$ is purely real as long as $|\alpha|<1$, i.e., when the effective Fermi velocity of NH Dirac fermions is purely real.

Although not directly relevant at this stage, it is worth noting that with a suitable definition of the spinor basis appropriate for monolayer graphene, a prototypical example of Dirac systems, the chiral symmetry breaking mass orders correspond to the charge-density-wave ($\Gamma_5$)~\cite{graphenemass:3} and two, namely the real ($\Gamma_3$) and imaginary ($\Gamma_4$), components of the Kekul\'e valence bond order~\cite{graphenemass:2, JaRoEx:2}, while the time-reversal symmetry breaking chiral scalar mass corresponds to the Haldane's quantum anomalous Hall insulator~\cite{graphenemass:4}. Furthermore, the same NH Dirac Hamiltonian in an appropriate Nambu-doubled spinor basis also describes the gapless surface states of three-dimensional NH topological insulators, devoid of any NH skin effect~\cite{salibNH}. In that case, $M=\Gamma_{12}$ corresponds to the surface magnetization in the $z$ direction, and $\Delta_1$ ($\Delta_2$) is the real (imaginary) component of an $s$-wave pairing~\cite{JaRoEx:1}. However, the following outcomes are insensitive to these details.

With the above construction for the NH Dirac operator in the presence of a composite mass order, next I proceed to consider a vortex-type defect therein for which 
\begin{equation}
{\boldsymbol \Delta} \to {\boldsymbol \Delta}(\vec{r})  = \Delta(r) \; \left( \cos(n \phi), \sin(n \phi) \right), 
\end{equation}
where $\Delta(r) \equiv |{\boldsymbol \Delta}(\vec{r})|$, $n \in {\mathbb Z}$ is the integer vorticity, $\phi$ is the azimuthal angle, and radial profile of the mass order is given by $\Delta(r \to 0)=0$ and $\Delta(r \to \infty)=\Delta_0$ (constant), otherwise arbitrary. In the presence of such vortex defect, the corresponding NH operator from Eq.~\eqref{eq:JRuniform} $H^{\rm NH}_{\rm Dir} (\vec{k},\alpha) \to H^{\rm NH}_{\rm Dir} (\vec{k} \to -i {\boldsymbol \nabla},\alpha,n)$ and then I look for the zero-energy right eigenstate in its spectrum by solving the following differential equation
\begin{align}~\label{eq:NHJackiwRossi}
&\bigg\{ \Gamma_1 \left[ 1 + \alpha \Gamma_{12} \right] (-i \partial_x) + \Gamma_2 \left[ 1 + \alpha \Gamma_{12} \right] (-i \partial_y) \nonumber \\ 
&+ \Delta(r) \left[ \Gamma_3 \cos(n \phi) + \Gamma_4 \sin(n\phi)\right] \bigg\} \ket{\Psi_{0}(r,\phi)}_R=0.
\end{align}
Finding the solution of the above equation becomes much more efficient by noticing the fact that $\{ H^{\rm NH}_{\rm Dir} (\vec{k} \to -i {\boldsymbol \nabla},\alpha,n), \Gamma_5 \}=0$. Hence, any mode that is bound at zero energy must then be an eigenstate of $\Gamma_5 ={\rm diag}. (1,-1,-1,1)$. Furthermore, I note that in the Hermitian limit ($\alpha=0$), $\{ H^{\rm NH}_{\rm Dir} (\vec{k} \to -i {\boldsymbol \nabla},0,n), {\mathcal A} \}=0$ where ${\mathcal A}=\Gamma_{23} {\mathcal K}$ is the generator of the antiunitary particle-hole symmetry~\cite{herbutlu}. Therefore, the zero-energy modes are also eigenstates of ${\mathcal A}$ in the Hermitian system. On the other hand, for any nontrivial $\alpha$, I find
\begin{eqnarray}~\label{eq:pseudoPHJR}
&&{\mathcal A} \; H^{\rm NH}_{\rm Dir} (\vec{k} \to -i {\boldsymbol \nabla},\alpha,n) {\mathcal A} 
= H^{\rm NH}_{\rm Dir} (\vec{k} \to -i {\boldsymbol \nabla},-\alpha,n) \nonumber \\
& \equiv & \left(  H^{\rm NH}_{\rm Dir} (\vec{k} \to -i {\boldsymbol \nabla},\alpha,n) \right)^{\dagger}
\end{eqnarray}
Therefore, the generator of the antiunitary particle-hole symmetry of Hermitian Jackiw-Rossi model connects the left and right zero-energy eigenmodes (when they exist) of its NH incarnation, yielding its pseudo-particle-hole symmetry. With these spectral symmetries of the Hermitian and NH Jackiw-Rossi operators in hand, I now proceed to find the explicit solutions for the zero-energy modes.

In what follows, I consider a vortex-type defect, characterized by an integer $n>0$. The solutions for an antivortex with $n<0$ can readily be obtained from the solutions below by noting that the operator (Hermitian or NH) for the antivortex is obtained after a unitary rotation of the operator with a vortex-type defect by $\Gamma_{45}$. Hence, zero mode solutions for the antivortex are also obtained after unitarily rotating the ones for a vortex-type defect by $\Gamma_{45}$, which I discuss next.

The zero-energy right eigenmodes localized near the core of the vortex whose amplitude falls off as $r \to \infty$ and square integrable with respect to its biorthogonal product with the left eigenmode near the origin as $r \to 0$ is then found to be of the following generic form
\begin{eqnarray}~\label{eq:zeromodeJRschematic}
\ket{\Psi_{0}(r,\phi)}_R= C \left( \begin{array}{c}
0 \\ 
v^R_1 (r,\phi) \\ 
u^R_2 (r,\phi) \\ 
0 
\end{array}
\right). 
\end{eqnarray}
Two functions $v^R_1 (r,\phi) \equiv v^R_1$ and $u^R_2 (r,\phi) \equiv u^R_2$ are obtained by substituting the following ansatz into Eq.~\eqref{eq:NHJackiwRossi}
\begin{eqnarray}
v^R_1 &=& \sqrt{\frac{1+\alpha}{1-\alpha}} \left[ f(r) e^{-im \phi} + g(r) e^{i(m+1-n)\phi}\right] e^{i\frac{\pi}{4}}, \nonumber \\
u^R_2 &=& \left[ f(r) e^{im \phi} + g(r) e^{-i(m+1-n)\phi}\right] e^{-i\frac{\pi}{4}}
\end{eqnarray}
The complete functional variation of $f(r)$ and $g(r)$ on the radial coordinate $r$ cannot be found in general. Upon substituting these ansatz into Eq.~\eqref{eq:NHJackiwRossi} I obtain their following asymptotic forms 
\begin{equation}
f( r \to 0) = c_0^f \; r^m 
\:\: \text{and} \:\:
g( r \to 0) = c_0^g \; r^{n-m-1}
\end{equation}
near the origin where $\Delta(r) \to 0$, while at a sufficiently large distance from the origin where $1/r \to 0$, these two functions take the form
\begin{equation}
f (r \to \infty) = g (r \to \infty) = c_\infty \; \exp\left( - \int^r_0 \frac{\Delta(t)}{\sqrt{1-\alpha^2}} \; dt \right),
\end{equation}
where $c_0^f$, $c_0^g$, and $c_\infty$ are three arbitrary constants that can be determined by matching the solutions at an intermediate value of $r$. For square normalizability of the solutions near the origin the integer values of $m$ (required for the single-valued solutions) are bounded within the range $-1/2 \leq m \leq n-1/2$, yielding $m=0,1,\cdots, n-1$. Therefore, there are a total of $n$ number of allowed values of $m$, leading to a total of $n$ number of zero-energy solutions, which is the celebrated Jackiw-Rossi index theorem, originally announced in Hermitian systems~\cite{jackiwrossi:1, jackiwrossi:2}. Also notice that for the zero mode solutions to be normalizable as $r \to \infty$, the NH parameter must satisfy $|\alpha| < 1$. Therefore, a NH planar massive Dirac system continues to foster $n$ number of normalizable zero-energy modes in the presence of a U(1) mass vortex, as long as the effective Fermi velocity of the collection of NH Dirac fermions remains purely real. When $|\alpha|>1$, the bound states I found otherwise for $|\alpha|<1$, become plane waves at zero energy. Existence of such zero-energy plane-wave modes in the presence of an underlying vortex-type defect solely results from the non-Hermiticity in the system.

When $n$ is an odd integer, there exists one special zero mode solution with $m=(n-1)/2$, for which a closed solution can be found for any arbitrary $r$. The corresponding right eigenvector is given by 
\begin{eqnarray}~\label{eq:specialzeroJR}
\ket{\Psi_{0}(r,\phi)}_R &=& C \left( \begin{array}{c}
0 \\ 
\sqrt{\frac{1+\alpha}{1-\alpha}} \\ 
-i \\ 
0 
\end{array}
\right) r^{\frac{n-1}{2}} \; \exp\left[{-i \left(\frac{n-1}{2}\right) \phi}\right] \nonumber \\  
&\times& \exp \left[ -\int^r_0 \frac{\Delta(t)}{\sqrt{1-\alpha^2}} \; dt\right].
\end{eqnarray}
This special zero-energy mode will play an important role when I search for the zero modes in the spectrum of a generalized NH Jackiw-Rossi operator, constructed by exploiting its pseudo-particle-hole symmetry, see Eq.~\eqref{eq:pseudoPHJR}, which I discuss next.

\subsection{Generalized NH Jackiw-Rossi operator}

The spectral symmetry of the Jackiw-Rossi Hamiltonian with respect to the antiunitary operator ${\mathcal A}=\Gamma_{23} {\mathcal K}$, allows one to introduce additional terms therein, leading to the \emph{generalized} Jackiw-Rossi Hamiltonian. When extended to the case of NH systems, where ${\mathcal A}$ generates a pseudo-particle-hole symmetry, the generalized NH Jackiw-Rossi operator takes the following form 
\allowdisplaybreaks[4] 
\begin{eqnarray}
H^{\rm Gen}_{\rm JR} &=& (1+\alpha \Gamma_{12}) \bigg[ \Gamma_1 (-i \partial_x -\Gamma_{34} A_x) + \Gamma_2 (-i \partial_y \nonumber \\
&-&\Gamma_{34} A_y) \bigg]
+ \Delta(r) \left[ \Gamma_3 \cos(n \phi) + \Gamma_4 \sin(n\phi)\right] \nonumber \\
&+& \mu \Gamma_{34} + h \Gamma_{12}. 
\end{eqnarray}
Notice that that the matrix operator $\Gamma_{34}$, appearing with the minimally coupled gauge field $\vec{A}$, causes the U(1) rotation between the real and imaginary components of the composite mass order, represented by $\Gamma_3$ and $\Gamma_4$, respectively. Hence, the presence of such gauge field is necessary for the stability of an isolated or deconfined vortex. The physical origin of such a gauge field and two other quantities, namely $\mu$ and $h$, depends on microscopic details. For example, in graphene-based Dirac system of spinless fermions the gauge field enters as the axial or chiral gauge field pointing in the opposite directions near two complementary valleys, and $\mu$ ($h$) corresponds to Haldane's quantum anomalous Hall insulator mass (chiral chemical potential). On the other hand, on the surface of three-dimensional topological insulators featuring a superconducting vortex, the gauge field results from regular electromagnetic vector potential, and $\mu$ ($h$) represents the $z$ component of surface magnetization (regular chemical potential). The forthcoming discussion in this Section on the existence of zero-energy modes in the spectrum of $H^{\rm Gen}_{\rm JR}$ will be guided by the existing results in the Hermitian system ($\alpha=0$)~\cite{goswamiroy}.

In Hermitian systems, it was shown that the spectrum of the generalized Jackiw-Rossi Hamiltonian features a \emph{single} zero-energy mode only when $n$ is an odd integer, for which the explicit solution smoothly deforms to the one shown in Eq.~\eqref{eq:specialzeroJR} in the limit when $\vec{A}, \mu, h \to 0$. By contrast, when $n$ is an even integer, there exists no mode at zero energy. Hence, in terms of the number of zero-energy modes in its spectrum, the generalized Jackiw-Rossi Hamiltonian displays a $Z_2$ index in contrast to the $Z$ index for the Jackiw-Rossi Hamiltonian, which I have discussed previously in this Section~\cite{goswamiroy}. In the presence of the gauge fields, it is not possible to find a closed form solution for the zero-energy mode for any arbitrary $r$. Thus, I present the right eigenmode for the zero-energy state when $\vec{A}=0$, but keeping $\mu$ and $h$ finite, which is explicitly given by 
\begin{eqnarray}
\ket{\Psi_{0}(r,\phi)}_R &=& C \left( \begin{array}{c}
u^R_1 \\ 
v^R_1 \\ 
(v^R_1)^\star \\ 
-(u^R_1)^\star 
\end{array}
\right) (r,\phi),
\end{eqnarray}
where 
\begin{eqnarray}~\label{eq:uvgenJR}
    u^R_1 &=& e^{-i \frac{\pi}{4}} \; e^{-i \left(\frac{n+1}{2}\right)\phi} \; g (r) \; \exp\left(-\int^r_0 \frac{\Delta(t)}{\sqrt{1-\alpha^2}} \; dt  \right), \nonumber \\
 \text{and} \;  
 v^R_1 &=& \sqrt{\frac{1+\alpha}{1-\alpha}} \; e^{-i \frac{\pi}{4}} \; e^{-i \left(\frac{n-1}{2}\right)\phi} \; f (r) \nonumber \\
&\times& \exp\left(-\int^r_0 \frac{\Delta(t)}{\sqrt{1-\alpha^2}} \; dt  \right).
\end{eqnarray}
The dependence of the zero modes on $\mu$ and $h$ is captured by the functions $f(r)$ and $g(r)$, which are respectively given by 
\begin{equation}~\label{eq:ffunction}
f(r) = \left\{
\begin{array}{rl}
C^{\mu>h}_1 \: J_{|\ell|} \left( \sqrt{\mu^2-h^2}r/\sqrt{1-\alpha^2} \right) & \text{if} \:\: \mu >h, \\
C^{h>\mu}_1 \: I_{|\ell|} \left( \sqrt{h^2-\mu^2}r/\sqrt{1-\alpha^2} \right) & \text{if} \:\: h>\mu.
\end{array} \right.
\end{equation}
and 
\begin{equation}~\label{eq:gfunction}
g(r) = \left\{
\begin{array}{rl}
C^{\mu>h}_2 \: J_{|m|} \left( \sqrt{\mu^2-h^2}r/\sqrt{1-\alpha^2} \right) & \text{if} \:\: \mu >h, \\
C^{h>\mu}_2 \: I_{|m|} \left( \sqrt{h^2-\mu^2}r/\sqrt{1-\alpha^2} \right) & \text{if} \:\: h>\mu.
\end{array} \right.
\end{equation}
Here, $\ell=-(n-1)/2$ and $m=-(n+1)/2$, $J_n$ and $I_n$ are the Bessel and modified Bessel functions of order $n$, respectively, and $C^{\mu>h}_{1,2}$ and $C^{h>\mu}_{1,2}$ are arbitrary constants that need to be determined from the overall normalization of the zero modes with respect to the biorthonormality condition, shown in Eq.~\eqref{eq:biorthonormality}.

When $\mu>h$, as the Bessel functions are well behaved near the origin and as $r \to \infty$, the normalization of the zero mode solutions is solely determined by the exponential factors in Eq.~\eqref{eq:uvgenJR}, thereby yielding normalizable solutions only if $|\alpha|<1$, i.e., when the effective Fermi velocity of the collection of NH Dirac fermions is real. Notice that when $h>\mu$, as the modified Bessel functions grow exponentially as $r \to \infty$, one can find normalizable solutions for zero modes if and only if $\Delta^2_0 + \mu^2 \geq h^2$ and when $|\alpha|<1$. The former condition is same as in Hermitian systems. Finally, I note that for $|\alpha|>1$, $J_n \leftrightarrow I_n$ and one cannot find any normalizable solution irrespective of the relative strength of $\mu$ and $h$.

Finally, I consider the effect of the orbital coupling of the external gauge field ($\vec{A}$) to NH Dirac fermions. I assume that the corresponding magnetic field (chiral or regular) is finite only within a distance $r \leq \xi$ and vanishes for $r>\xi$, where $\xi$ determines the core size of the vortex~\cite{herbutlu, goswamiroy, melikyantesanovic, juricictesanovic}. I work in the symmetric gauge in which $\vec{A}=(A_r,A_\phi)$ with $A_r=0$ and  
\begin{equation}
A_\phi= \begin{cases}
1/(2r) & \text{when} \:\:\: r > \xi,\\
r^2/(2 \xi) & \text{when} \:\:\: r \leq \xi.
\end{cases}
\end{equation}
With such a profile of the magnetic field (either chiral or regular, depending on the spinor basis), the system supports a single vortex with vorticity $n=1$ and from now on I focus on such single vortex with unit vorticity. Outside the vortex core at large distance, I find 
\begin{equation*}
f(r) = \left\{
\begin{array}{rl}
\sum_{j=\pm 1} C^{\mu>h}_{j} \: J_{j /2} \left( \sqrt{\frac{\mu^2-h^2}{1-\alpha^2}} \; r \right) & \text{if} \:\: \mu >h, \\
\sum_{j=\pm 1} C^{h>\mu}_{j} \: I_{j /2} \left( \sqrt{\frac{\mu^2-h^2}{1-\alpha^2}} \; r \right) & \text{if} \:\: h>\mu.
\end{array} \right.
\end{equation*}
and 
\begin{equation*}
g(r) = \left\{
\begin{array}{rl}
\sum_{j=\pm 1} C^{\mu>h}_{j} \: j \; J_{j /2} \left( \sqrt{\frac{\mu^2-h^2}{1-\alpha^2}} \; r \right) & \text{if} \:\: \mu >h, \\
\sum_{j=\pm 1} C^{h>\mu}_{j} \: j \; I_{j /2} \left( \sqrt{\frac{\mu^2-h^2}{1-\alpha^2}} \; r \right) & \text{if} \:\: h>\mu.
\end{array} \right.
\end{equation*}
On the other hand, inside the vortex core the functions $f(r)$ and $g(r)$ take the form shown in Eqs.~\eqref{eq:ffunction} and~\eqref{eq:gfunction}, respectively, with $n=1$ and $\mu^2-1/(2 \xi^2) \to \mu^2$ therein. Even in this case, I find that normalizable zero energy modes exist only when $|\alpha|<1$, i.e., when the Fermi velocity of the collection of NH Dirac fermions is real.

\section{Summary and discussion}~\label{sec:discussion}

To summarize, I here extend the jurisdiction of three prominent index theorems for quasirelativistic Dirac fermions to open or NH systems, for which the NH Dirac operators besides featuring the Lorentz invariance also display a purely real energy eigenvalue spectrum over an extended NH parameter regime. Such NH planar Dirac systems, when placed in external magnetic fields continue to host robust zero-energy modes, the number of which is equal to the number of magnetic flux quanta enclosed by the system, irrespective of the strength of the NH perturbation and the profile of the magnetic field in the system. On the other hand, when massive NH Dirac fermions sustain a topologically nontrivial texture of the mass order in the real space, such as a domain wall in one dimension and vortices in two dimensions, the system still honors the Jackiw-Rebbi and Jackiw-Rossi index theorems, respectively, as in the Hermitian systems, but only when the effective Fermi velocity for NH Dirac fermions is purely real. In these cases, on the other hand, when the Fermi velocity becomes imaginary, the zero-energy modes become plane waves, a phenomenon exclusively occurring in NH systems. In future, it will be worthwhile finding appropriate interacting models that can sustain the desired ``commuting class mass'' orders in NH Dirac system in which one can then study the imprints of topological defects~\cite{NHDirac:1, NHDirac:2, NHDirac:3, NHDirac:4, NHDirac:5, NHinteraction}.

I also notice that for $|\alpha|<1$ and even more interestingly when $|\alpha|>1$ (yielding an imaginary Fermi velocity), the vortex core with at least $|n|=1$ continues to support a single normalizable zero-energy mode when the system is subject to regular magnetic fields (not the deconfining one) of arbitrary spatial profile, the effects of which are incorporated by taking $-i\partial _j-A_j$ for $j=x$ and $y$ (minimal coupling with the magnetic vector potential $\vec{A}$) in Eq.~\eqref{eq:NHJackiwRossi}.  The resulting NH gauged-Dirac operator enjoys the unitary particle-hole symmetry $\{ H^{\rm NH}_{\rm Dir} (\vec{k} \to -i {\boldsymbol \nabla}-\vec{A},\alpha,n), \Gamma_5 \}=0$, which protects the zero modes of the schematic form shown in Eq.~\eqref{eq:zeromodeJRschematic}. This conclusion follows from the one in Hermitian systems~\cite{realBIndex:1, realBIndex:2, realBIndex:3}, with the solutions expressed in terms of parabolic cylindrical functions~\cite{realBIndex:4}, when the magnetic field is uniform. The outcomes can be justified in the following way. Notice that in the presence of magnetic fields ($\vec{B}$) in a system with a mass vortex, there are two length scales, namely the magnetic length ($\ell_B$) and a length scale due to the mass order $\ell_\Delta=(\Delta_0/\sqrt{1-\alpha^2})^{-1}$ that together determine the localization of the zero-energy mode bound to such defect. For $|\alpha|<1$, both $\ell_B$ and $\ell_\Delta$ are real, and the solutions of the zero-energy modes in NH systems are smoothly connected to the one in Hermitian setups. For $|\alpha|>1$, only $\ell_B$ is real, which then solely determines the localization of the normalizable zero-energy modes. Only for the sake of brevity do I not show here the explicit solutions of such a mode.

These findings should be of far reaching consequences in open or NH quasirelativistic systems. For example, existence of zero-energy modes in the presence of external magnetic fields of arbitrary shape strongly suggests that the magnetic catalysis mechanism for the nucleation of chiral symmetry breaking mass orders should remain operative in weakly interacting NH planar Dirac systems~\cite{aharonovcasher:5, catalysis:1, catalysis:2}. Dynamic symmetry breaking following this mechanism causes insulation in the system at half-filling and gives rise to the notion of competing orders within the zero-energy manifold. Zero-energy bound states, localized near the core of topological defects of mass orders, can give rise to charge fractionalization (for charged Dirac fermions) and localized Majorana modes (for neutral Bogoliubov Dirac quasiparticles) in open quasirelativistic systems. The localized zero-energy modes near the core of topological defects of mass orders can also foster additional and distinct local mass order parameters~\cite{JaRoEx:9, JaRoEx:10, JaRoEx:11}. Therefore, when such defects proliferate in the system, local mass orders can acquire a global coherence, leading the notion of deconfined quantum phase transition, which thus far has been discussed exclusively in closed or Hermitian systems~\cite{deconfined:1, deconfined:2, deconfined:3, deconfined:4}. The present discussion, therefore, opens an uncharted territory of such continuous deconfined phase transition between two distinct ordered phases via the proliferation of real space topological defects in open or NH systems. The index theorems and the existence of zero-energy modes in NH systems are also germane in the context of topological phases of matter~\cite{salibNH}.

The findings related to the Jackiw-Rossi index theorem extend directly to the situation with a $z$-directional line vortex in three-dimensional NH massive Dirac and Weyl systems~\cite{goswamiroy, nishida}, suggesting the presence of dispersive plane-wave modes along the vortex line that are localized around the vortex core as long as the effective Fermi velocity of NH Dirac fermions is real. In this case, the celebrated Callan-Harvey mechanism becomes operative due to the presence of one-dimensional modes along the vortex core~\cite{callanharvey}, leading to a flow of nondissipative quantized current, given by $j_z= n e^2 E_z/(2\pi)$, when an electric field ($E_z$) is applied in the $z$ direction, where $n$ is the number of zero-energy localized modes bound to the vortex core. Such a current will be supplied radially into the vortex core which can be captured by an axion electrodynamics term. A detailed derivation of this phenomenon is, however, left for a future investigation. The present discussion can also be extended to string-type configurations for a general fermionic mass matrix for neutrinos in an extended standard model~\cite{Stojkovic:2}, but in a NH formalism.

The simplicity of the construction of the NH Dirac operator, see Sec.~\ref{subsec:construction}, makes the theoretical predictions from this work testable in tabletop experiments. For example, a collection of Hermitian massless Dirac fermions can be realized on graphene's honeycomb lattice, which can be captured from the nearest-neighbor tight-binding Hamiltonian. In a two-component spinor basis $\Psi^\top_{\bf k}=(c_A, c_B)({\bf k})$, where $c_A({\bf k})$ and $c_B({\bf k})$ are the fermionic annihilation operators on the sites of two interpenetrating triangular sublattices $A$ and $B$, respectively, of the honeycomb lattice such a tight-binding model with hopping amplitude $t_0$ leads to the following Bloch Hamiltonian 
\begin{equation}~\label{eq:TBHoneycomb}
h^{\rm lat}_0({\bf k})= t_0 \; \left( \begin{array}{cc}
0 & f({\bf k}) \\
f^\star({\bf k}) & 0
\end{array} \right), 
\end{equation}
where $f({\bf k})=\exp[i {\bf k} \cdot {\bf b}_1] + \exp[i {\bf k} \cdot {\bf b}_2] + \exp[i {\bf k} \cdot {\bf b}_3]$, and ${\bf b}_1=(1/\sqrt{3},1)a/2$, ${\bf b}_2=(1/\sqrt{3},-1)a/2$, and ${\bf b}_3=(-1/\sqrt{3},0)a$ are the nearest-neighbor vectors. Here, $a$ is the lattice spacing and ``$\star$'' denotes the complex conjugation. The above Bloch Hamiltonian fosters massless Dirac fermions around six corners of the hexagonal Brillouin zone. With a specific choice of $M={\rm diag.}(1,-1)$ (yielding a staggered potential between two sublattices of the honeycomb lattice~\cite{graphenemass:3}) in Eq.~\eqref{eq:NHDirac}, I then realize a collection of NH massless Dirac fermions, where the non-Hermiticity results from the imbalance between the hopping amplitudes in the opposite directions between two sublattices. Such a platform yields the ideal setup where NH generalization of the Aharonov-Casher index theorem can be tested at least on optical honeycomb lattices~\cite{opticalgraphene:1} on which the orbital effects of magnetic field can be engineered by synthetic gauge fields~\cite{syntheticgauge:1, syntheticgauge:2} and hopping imbalance yielding non-Hermiticity~\cite{NHoptLatt:1, NHoptLatt:2} can in principle be emulated.

In the same setup the Haldane's quantum anomalous Hall insulator is represented by a mass matrix $M={\rm diag.}(1,-1) g(\vec{k})$, where $g(\vec{k})=\sin(i {\bf k} \cdot {\bf a}_1) + \sin(i {\bf k} \cdot {\bf a}_2) + \sin(i {\bf k} \cdot [{\bf a}_2 - {\bf a}_1])$ and ${\bf a}_{1}=(\sqrt{3},-1)a/2$ and ${\bf a}_{2}=(0,1)a$ are Bravais lattice vectors of the triangular lattice~\cite{graphenemass:4}. With such a choice of $M$, the corresponding NH Dirac operator from Eq.~\eqref{eq:NHDirac} also yields an imbalance of the hopping amplitudes between the nearest-neighbor sites in the opposite directions. Since Haldane's anomalous Hall insulator order has already been engineered on optical honeycomb lattice~\cite{haldaneoplatt}, it is natural to expect that such an intrasublattice circulating current pattern can be utilized to engineer the corresponding NH Dirac operator therein following the general protocol from Sec.~\ref{subsec:construction}. On such a setup, a Y junction of Kekul\'e valence bond order [yielding lattice regularization of a U(1) mass vortex] can bind localized zero-energy modes to feature charge fractionalization for NH Dirac fermions. Furthermore, three-dimensional Weyl fermions have also been realized on optical lattices recently~\cite{WeylOptLatt}, constituting the ideal stage on which an extension of the Callan-Harvey mechanism, for example, in a NH Weyl system can be showcased, as conjectured from the present discussion, especially given the recent success in engineering non-Hermiticity in cold atomic systems~\cite{NHoptLatt:2}.

Besides optical honeycomb lattices, artificial honeycomb lattices also constitute another promising platform where the predicted zero modes bound to a mass vortex causing charge fractionalization can be observed since on such systems the Kekul\'e valence bond order has already been realized~\cite{artficialgrapheneMott:1, artficialgrapheneMott:2}. Altogether, the Lorentz invariant NH Dirac theory constitutes an ideal platform where the footprints of various celebrated index theorems in open quasirelativistic systems can be tested theoretically and their tabletop experimental verification should be within the reach of currently accessible facilities to engineer desired NH quasirelativistic lattices with balanced gain and loss (required for all-real eigenvalues of a NH operator).

Extending the index theorems to non-Lorentz invariant NH systems should be a interesting generalization of the current pursuit. Note that Lorentz symmetry can be violated in numerous ways and encompassing all of them certainly goes beyond the scope of the present discussion. Nonetheless, here I discuss one such case in two dimensions ($d=2$) where the conclusions from the present work can be extended straightforwardly to predict the existence of topologically robust zero-energy states, following the existing results in Hermitian systems and the general principle of constructing the corresponding NH operator from Sec.~\ref{subsec:construction}. For example, one can consider the following generalization of the Dirac Hamiltonian from Eq.~\eqref{eq:DiracHamil} describing an $s$th order touching of the valence and the conduction bands for which the Hamiltonian reads as 
\begin{equation}
H_s (\vec{k})= \Lambda_s \sum_{j=1,2} \Gamma_j \; d^s_j (\vec{k})
\end{equation}
with $\vec{d}^s(\vec{k})= (d^s_1,d^s_2)(\vec{k})=|k|^s (\cos(s \phi_{\vec{k}}), \sin(s \phi_{\vec{k}}))$, where $\phi_{\vec{k}}=\tan^{-1}(k_y/k_x)$ and $\Lambda_s$ bears the appropriate dimension such that $H_s$ carries the dimension of energy. For example, when $s=2$, the parameter $\Lambda_s$ has the dimension of inverse mass. Throughout this work, I considered only Lorentz symmetric systems with $s=1$ for which $\Lambda_s \equiv v_{_{\rm H}}$. Due to the nonlinear energy momentum relation $E(\vec{k})= \pm \Lambda_s \; |\vec{k}|^s$, the Lorentz symmetry gets violated for any integer $s>1$. Nonetheless, the entire construction of the corresponding NH operator now follows straightforwardly from Sec.~\ref{subsec:construction} by taking $H_{\rm Dir}(\vec{k}) \to H_s(\vec{k})$ therein, as there exist a plethora of mass order matrices $M$ that satisfy $\{ H_s(\vec{k}), M \}=0$.

The Aharonov-Cashed index theorem in such a Hermitian system guarantees the existence of $s N$ number of topologically robust zero-energy modes whose spatial localization is solely determined by the profile of the external (uniform or inhomogeneous) magnetic fields ($\Lambda_s$ playing no role)~\cite{aharonovcasher:4}, where $N$ is the number of magnetic flux quanta enclosed by the system. As the NH counterpart features the same unitary particle-hole symmetry (generated by $\sigma_3$), I expect the corresponding operator to support $s N$ number of normalizable zero-energy modes, irrespective of the strength of the non-Hermiticity $\alpha$, as shown explicitly for $s=1$ here. The Jackiw-Rossi index theorem in such Hermitian models with an underlying mass vortex of vorticity one yields $s$ number of normalizable zero-energy modes~\cite{quadraticIndexJR:1, quadraticIndexJR:2}, which are expected to survive (due to the unitary particle-hole symmetry, generated by $\Gamma_5$) in NH systems as long as $|\alpha|<1$. Existence of such $s$ number of zero-energy modes can be responsible for charge $se$ superconductivity~\cite{quadraticIndexJR:3}, which should be also operative in the analogous NH setups. Explicit demonstrations of these outcomes are, however, left for future investigations. For other index theorems in Lorentz-violating NH systems, see Ref.~\cite{NHtopologyIndex}.

\acknowledgments

The author is thankful to Christopher A.\ Leong and Vladimir Juri\v ci\' c for critical reading of the manuscript. B.R.\ was supported by NSF CAREER Grant No.~DMR-2238679 (USA) and thanks ANRF, India, for support through the Vajra scheme
VJR/2022/000022.

\section*{Data Availability}
No data were created or analyzed in this study.

\end{document}